\documentclass[aps,prd,showpacs,twocolumn]{revtex4}
\usepackage{amssymb}
\usepackage{amsmath}

\def\be{\begin{equation}}
\def\ee{\end{equation}}
\def\bea{\begin{eqnarray}}
\def\eea{\end{eqnarray}}

\begin{document}

\title{The virial theorem and the dark matter problem in  hybrid metric-Palatini gravity}

\author{Salvatore Capozziello$^{1.2}$}\email{capozzie@na.infn.it}
\author{Tiberiu Harko$^3$}\email{t.harko@ucl.ac.uk}
\author{Tomi S. Koivisto$^{4}$}\email{tomi.koivisto@fys.uio.no}
\author{Francisco S.N.~Lobo$^{5}$}\email{flobo@cii.fc.ul.pt}
\author{Gonzalo J. Olmo$^{6}$}\email{gonzalo.olmo@csic.es}

\affiliation{$^1$Dipartimento di Scienze Fisiche, Universit\`{a} di Napoli "Federico II", Napoli, Italy}
\affiliation {$^2$INFN Sez. di Napoli, Compl. Univ. di Monte S. Angelo, Edificio G, Via Cinthia, I-80126, Napoli, Italy}
\affiliation{$^3$Department of Mathematics, University College London, Gower Street, London WC1E 6BT, United Kingdom}
\affiliation{$^{4}$  Institute of Theoretical Astrophysics, University of
  Oslo, P.O.\ Box 1029 Blindern, N-0315 Oslo, Norway}
\affiliation{$^5$Centro de Astronomia e Astrof\'{\i}sica da Universidade de Lisboa, Campo Grande, Ed. C8 1749-016 Lisboa, Portugal}
\affiliation{$^6$Departamento de F\'{i}sica Te\'{o}rica and IFIC, Centro Mixto Universidad de
Valencia - CSIC. Universidad de Valencia, Burjassot-46100, Valencia, Spain}

\date{\today}

\begin{abstract}

Hybrid metric-Palatini gravity is a recently proposed theory, consisting of the superposition of the metric Einstein-Hilbert Lagrangian with an $f(\cal R)$ term constructed \`{a} la Palatini. The theory predicts the existence of a long-range scalar field, which passes the Solar System observational constraints, even if the scalar field is very light, and modifies the cosmological and galactic dynamics.
Thus, the theory opens new possibilities to approach, in the same theoretical framework, the problems of both dark energy and dark matter.
In this work, we consider the generalized virial theorem in the scalar-tensor representation of the hybrid metric-Palatini gravity. More specifically, taking into account the relativistic collisionless Boltzmann equation, we show that the supplementary geometric terms in the gravitational field equations provide an effective contribution to the gravitational potential energy. We show that the total virial mass is proportional to the effective mass associated with the new terms generated by the effective scalar field, and the baryonic mass. 
In addition to this, we also consider astrophysical applications of the model and show that the model predicts that the mass associated to the scalar field and its effects extend beyond the virial radius of the clusters of galaxies. In the context of the galaxy cluster velocity dispersion profiles predicted by the hybrid metric-Palatini model, the generalized virial theorem can be an efficient tool in observationally testing the viability of this class of generalized gravity models. 
\end{abstract}

\pacs{04.50.+h, 04.20.Jb, 04.20.Cv, 95.35.+d}

\maketitle

\section{Introduction}
\label{sec:a}

Modern astrophysics and cosmology are facing two major intriguing challenges, namely, the dark energy and the dark matter problems, respectively. Two important astronomical observations, the flat rotations curves of galaxies, and the virial mass discrepancy in clusters of galaxies led to the necessity of inferring the existence of a special form of matter, called dark matter, which interacts with baryonic matter only gravitationally, and whose presence can explain the dynamical behavior of test particles at the galactic and extra-galactic scale. On the galactic or intra-galactic scale, the astronomical observations show a linear mass increase even in regions where very little baryonic matter can be detected, with the rotational velocities attaining an approximately constant value, $v_{tg\infty} \sim 200-300 {\rm km}/{\rm s}$, within a distance $r$ from the center of the galaxy~\cite{Bi87}.

The determination of the total mass that is gravitationally bounded in clusters of galaxies is important for our understanding of the nature and evolution of structures on cosmological scales. The gravitational masses of clusters of galaxies are estimated by assuming a hydrostatic equilibrium of both the hot intra-cluster gas and of the galaxies with the binding cluster potential. Therefore the total mass of a cluster of galaxies can be estimated in two ways. In the first method, by taking into account the dynamical motions of the member galaxies of the cluster, and with the application of the virial theorem,  one obtains an estimate $M_{V}$ for the mass of the cluster. Secondly, the total baryonic mass $M_B$ can be determined  by adding the mass of each individual galaxy member of the cluster. The mass discrepancy at the galactic cluster level arises as observations show that $M_{V}$ is much greater than $M_B$, with typical values of $M_{V}/M \sim 20-30$~\cite{Bi87}.

These important astrophysical observations  are usually explained by postulating the existence of a new form of matter, called dark matter, assumed to be a cold pressure-less fluid extended in a spherically symmetric halo around the galaxies and clusters (for a recent review on the dark matter properties see \cite{Sal}). Many candidates for dark matter particles have been proposed,  the most popular ones being the Weakly Interacting Massive Particles (WIMPs)~\cite{OvWe04}. The interaction cross section of WIMPs with standard baryonic matter is assumed to be extremely small, but non-zero, and therefore these particles may be directly observable experimentally. Recently, a pressure-less and non-comoving two-fluid dark matter model has also been analysed, in which dark matter is represented as a two-component fluid thermodynamic system, without interaction between the constituent particles of different species \cite{Harko:2011nu}.

However, despite the extensive experimental and observational investigations, up to now \textit{non-gravitational} evidence for dark matter is still lacking. Moreover, major accelerator and reactor experiments, like LHC, did not bring convincing evidence for the existence of new physics beyond the standard model, on which the dark matter hypothesis is based upon. Therefore, the possibility that Einstein's (and Newton's) theory cannot describe  gravitational phenomena at the scale of galaxies and clusters of galaxies must not be excluded \textit{a priori}. Several theoretical approaches, in which ``dark matter'' can be understood as a modification of the gravitational laws at large scales have been extensively proposed in the literature~\cite{dark}.

The virial theorem, which gives a simple relation between the kinetic and potential energy of a system of particles,  plays an important role in astrophysics and cosmology \cite{Col}. By observing the velocities of test particles and by assuming hydrostatic equilibrium, with the use of the virial theorem, one can obtain the mean density of astrophysical objects such as galaxies, clusters, or super clusters. An important application of  the virial theorem is the determination of the total mass of the clusters of galaxies. The virial theorem is also a powerful tool for the study of the stability of gravitationally bounded objects. The extension of the Newtonian virial theorem to the general relativistic case has led to several versions of the virial theorem~\cite{vir}, including the effect of a cosmological constant~\cite{Ja70,No}, the generalization to brane world models~\cite{HaCh07},  $f(R)$ gravity  \cite{BoHaLo08}, DGP brane models \cite{Shah}, and to Palatini $f(R)$ models \cite{Sep}.

In this context,  a novel approach to modified theories of gravity was proposed \cite{fX}, that consists of adding to the Einstein-Hilbert Lagrangian a $f(R)$ term constructed within the framework of the  Palatini formalism. Using the respective dynamically equivalent scalar-tensor representation, even if the scalar field is very light, the theory  passes the Solar System observational constraints. Therefore the long-range scalar field is able to modify the cosmological and galactic dynamics, but leaves the Solar System unaffected. The absence of instabilities in perturbations was also verified, and explicit models, which are consistent with local tests and lead to the late-time cosmic acceleration  were also found.

The cosmological applications of the hybrid metric-Palatini gravitational theory were investigated in \cite{fX1}. Criteria to obtain cosmic acceleration were analyzed, and the field equations were formulated  as a dynamical system. Several classes of  cosmological solutions, depending on the functional form of the effective scalar field potential, describing both accelerating and decelerating universes, were explicitly obtained. Furthermore, the cosmological perturbation equations were derived and applied to uncover the nature of the propagating scalar degree of freedom and the signatures of  these models predicted in the large-scale structure.
In addition to this, the hybrid metric-Palatini theory was considered in wormhole physics \cite{fX2}.
The general conditions for wormhole solutions according to the null energy conditions at the throat in the hybrid metric-Palatini gravitational theory were presented in \cite{fX2}. Several wormhole type solutions were also obtained and analyzed. In the first solution, the redshift function and the scalar field were specified, and the potential was chosen so that the modified Klein-Gordon equation can be simplified. This solution is not asymptotically flat and needs to be matched to a vacuum solution. In the second example, by adequately specifying the metric functions and choosing the scalar field, one can find an asymptotically flat spacetime.

The purpose of the present paper is to check  if the effective matter term induced by the equivalent scalar field in the gravitational field equations
 can explain the dark matter effect in clusters of galaxies. In order to find an answer to this question we investigate the virial theorem in the framework of hybrid metric-Palatini gravity. Using the collisionless Boltzmann equation in the modified Einstein field equations  we derive a generalized virial equality for the hybrid model. The generalized virial theorem  takes into account the presence of the supplementary scalar field related terms,  which do appear due to the modification of the gravitational action.
These supplementary geometric terms, and their scalar field equivalent,  give an effective contribution to the gravitational potential energy, with the total virial mass  being given by the sum of the effective mass associated to the new scalar field related terms, and the baryonic mass, respectively. Therefore the new scalar field  may account for the  virial theorem mass discrepancy in clusters of galaxies \cite{Bi87}. The gravitational field equations of the hybrid metric-Palatini gravitational model together with the virial theorem also allow to obtain the metric inside the cluster of galaxies  in a simple form, as functions of physical parameters that can be fully determined from astrophysical observations, like, for example, the temperature of the intra-cluster gas and the radius and central density of the cluster core.  Therefore the generalized virial theorem in hybrid metric-Palatini gravity can be an efficient tool in observationally testing the viability of this class of generalized gravity models.

The present paper is organized as follows. In Section \ref{sec:b2}, the gravitational field equations for spherically symmetric galactic clusters in the scalar-tensor version of hybrid metric-Palatini gravity and the relativistic Boltzmann equation are explored. In Section~\ref{sec:c}, the generalized virial theorem in hybrid metric-Palatini gravity is derived.  Astrophysical applications of the virial theorem are explored in Section~\ref{sec:d}. In particular, predictions of the geometric mass and geometric radius from galactic cluster observations are presented, and the behavior of the galaxy cluster velocity dispersion in hybrid metric-Palatini gravity models is considered.  Finally, in Section~\ref{sec:e}, we discuss and conclude our results.

\section{Galactic clusters: Basic formalism and field equations for a system of identical and collisionless
point particles }\label{sec:b2}

Astronomical and astrophysical observations have proved that galaxies tend to concentrate in
larger structures, called clusters of galaxies, with  total masses ranging from $10^{13} M_{\odot }$ for groups up to a few $10^{15} M_{\odot}$ for very large systems.  The cluster morphology is usually
dominated by a regular, centrally peaked main component~\cite{ReBo02,Ar05}.
Since clusters are considered to be ``dark matter'' dominated astronomical systems, their formation and evolution is controlled  by the gravitational force. The mass function of the
clusters is determined by the initial conditions of the mass distribution
set in the early universe~\cite{Sch01}.

In the present Section we introduce the variational formulation and the field equations of the hybrid metric-Palatini gravity theory, and write down the field equations for a system of identical and collisionless spherically symmetric distributed point particles.

\subsection{Hybrid metric-Palatini gravity: The formalism}

The hybrid metric-Palatini gravity model can be formulated in a scalar-tensor representation by starting from the action \cite{fX},
\begin{equation} \label{scalar2}
S= \frac{1}{2\kappa^2}\int d^4 x \sqrt{-g} \left[ (1+\phi)R +\frac{3}{2\phi}\nabla _\mu \phi \nabla ^\mu \phi
-V(\phi)  \right] +S_m \,,
\end{equation}
where $S_m$ is the matter action, and $\kappa ^2=8\pi G/c^3$, respectively. $V(\phi )$ is the scalar field potential.  Note that the gravitational theory given by Eq.~(\ref{scalar2}) is equivalent  with the purely metric Brans-Dicke-like action, with the Brans-Dicke  parameter $w=-3/2$, but with a different coupling to matter.

The variation of this action with respect to the metric tensor gives the field equations
\begin{eqnarray}\label{einstein_phi}
(1+\phi)G_{\mu\nu}&=&\kappa^2T_{\mu\nu} + \nabla_\mu\nabla_\nu\phi - \Box\phi\/g_{\mu\nu}
\nonumber\\
&& \hspace{-1.5cm} -\frac{3}{2\phi}\nabla_\mu\phi
\nabla_\nu\phi + \frac{3}{4\phi}\nabla_\lambda\phi\nabla^\lambda\phi g_{\mu \nu}- \frac{1}{2}Vg_{\mu\nu},
\end{eqnarray}
where $T_{\mu \nu}$ is the matter energy-momentum tensor.
Varying the action with respect to the scalar field we obtain
\be\label{variation_phi}
R - \frac{3}{\phi}\Box\phi +\frac{3}{2\phi^2}\nabla _\mu \phi \nabla ^\mu \phi - \frac{dV}{d\phi} =0
\ee
Moreover, one can show that the identity
\begin{equation}\label{eq:phi(X)}
2V-\phi \frac{dV}{d\phi}=\kappa^2T+R \ ,
\end{equation}
where $T=T_{\mu }^{\mu}$, also holds, and that the scalar field $\phi$ is
governed by the second-order evolution equation
\begin{equation}\label{eq:evol-phi}
-\Box\phi+\frac{1}{2\phi}\nabla _\mu \phi \nabla ^\mu
\phi+\frac{1}{3}\phi\left[2V-(1+\phi)\frac{dV}{d\phi}\right]=\frac{\phi\kappa^2}{3}T\,,
\end{equation}
which is an effective Klein-Gordon equation.

\subsection{Field equations for a system of identical and collisionless
point particles}

In order to derive the generalization of the relativistic virial theorem for
galaxy clusters in the hybrid metric-Palatini gravitational models we need, as a first step, to
obtain the gravitational field equations for a static and spherically
symmetric distribution of matter. To thus effect, consider a self-gravitating system of identical, collisionless point particles in random motion. To
obtain the basic field equations we will use the scalar-tensor
representation of hybrid metric-Palatini gravity, which allows a clear physical
interpretation of the model.

We assume that the geometry of the cluster can be described by a time-oriented Lorentzian four-dimensional space-time
manifold $\mathcal{M}$. The metric of an isolated  spherically symmetric
cluster is given by
\be
ds^{2}=-e^{\nu \left( r\right) }dt^{2}+e^{\lambda \left( r\right)
}dr^{2}+r^{2}\left( d\theta ^{2}+\sin^{2}\negmedspace\theta d\varphi
^{2}\right).  \label{line}
\ee
The galaxies in the cluster are considered identical and collisionless point
particles, and their space-time distribution  is described by a distribution function $f_B$. The latter function obeys the general relativistic Boltzmann equation, which will be presented in the next Section.

Thus, the energy-momentum tensor of matter is determined by the distribution function $f_B$, and is given by~\cite{Li66}
\be
T_{\mu \nu }=\int f_B\, m\, u_{\mu }u_{\nu }\;du,
\ee
where $m$ is the mass of the particle (galaxy), $u_{\mu }=\left(u_t,u_r,u_{\theta},u_{\varphi}\right)$ is the
four-velocity of the galaxy, with $u_t$ denoting the temporal component, and $du =du_{r}du_{\theta }du_{\varphi }/u_{t}$
is the invariant volume element of the velocity space, respectively. The
energy-momentum tensor $T_{\mu \nu }$ of the matter in a cluster of galaxies can be
represented in terms of an effective density $\rho_{\mathrm{eff}}$ and of two
effective anisotropic pressures, the radial $p_{\mathrm{eff}}^{(r)}$ and the
tangential $p_{\mathrm{eff}}^{(\perp)}$ pressures, respectively, given by
\be
\rho_{\mathrm{eff}} = \rho \left\langle u_{t}^{2}\right\rangle, p_{%
\mathrm{eff}}^{(r)}=\rho \left\langle u_{r}^{2}\right\rangle,
p_{\mathrm{eff}}^{(\perp)} = \rho \left\langle u_{\theta
}^{2}\right\rangle= \rho \left\langle u_{\varphi }^{2}\right\rangle,
\ee
where $\rho $ is the mass density of the ordinary baryonic matter, and $\left\langle u_{i}^{2}\right\rangle $, $i=t,r,\theta ,\varphi $ is the average
value of $u_{i}^{2}$, $i=t,r,\theta ,\varphi $, the square of the components of the four-velocity~\cite{Ja70}.

By using this form of the energy-momentum tensor, the gravitational field
equations describing a cluster of galaxies in hybrid metric-Palatini gravity take the form
\begin{eqnarray} \label{f1}
-e^{-\lambda }\left( \frac{1}{r^{2}}-\frac{\lambda ^{\prime }}{r}\right) +
\frac{1}{r^{2}}=8\pi \frac{G}{1+\phi }\rho \left\langle u_{t}^{2}\right\rangle
 \nonumber \\
-\frac{1}{2\left(1+\phi \right) }V\left( \phi \right) +\frac{1}{1+\phi }\left( \nabla
_{t}\nabla ^{t}-\square \right) \phi ,
 \nonumber \\
-\frac{3}{2}\frac{1}{\phi (1+\phi )}\nabla _t \phi \nabla ^t\phi +\frac{3}{4}\frac{1}{\phi \left(1+\phi \right)}\nabla _{\lambda }\phi \nabla ^{\lambda }\phi ,
\end{eqnarray}
\begin{eqnarray}\label{f2}
e^{-\lambda }\left( \frac{\nu ^{\prime }}{r}+\frac{1}{r^{2}}\right) -\frac{1
}{r^{2}}=8\pi \frac{G}{1+\phi }\rho \left\langle u_{r}^{2}\right\rangle
  \nonumber  \\
+\frac{1}{2\left(1+\phi \right) }V\left( \phi \right) -\frac{1}{1+\phi }\left( \nabla
_{r}\nabla ^{r}-\square \right) \phi ,
   \nonumber \\
+\frac{3}{2}\frac{1}{\phi (1+\phi )}\nabla _r \phi \nabla ^r\phi -\frac{3}{4}\frac{1}{\phi \left(1+\phi \right)}\nabla _{\lambda }\phi \nabla ^{\lambda }\phi ,
\end{eqnarray}
\begin{eqnarray}
\frac{1}{2}e^{-\lambda }\left( \nu ^{\prime \prime }+\frac{\nu ^{\prime 2}}{%
2}+\frac{\nu ^{\prime }-\lambda ^{\prime }}{r}-\frac{\nu ^{\prime }\lambda
^{\prime }}{2}\right)  =8\pi \frac{G}{1+\phi }\rho \left\langle u_{\theta }^{2}\right\rangle \notag
\\
 +\frac{1}{2\left(1+\phi \right) }V\left( \phi \right) -\frac{1}{1+\phi }\left( \nabla \notag
_{\theta }\nabla ^{\theta }-\square \right) \phi
\\
 +\frac{3}{2}\frac{1}{\phi (1+\phi )}\nabla _\theta  \phi \nabla ^\theta \phi -\frac{3}{4}\frac{1}{\phi \left(1+\phi \right)}\nabla _{\lambda }\phi \nabla ^{\lambda }\phi  ,
\\
=8\pi \frac{G}{1+\phi }\rho \left\langle u_{\varphi }^{2}\right\rangle
   \nonumber \\
-
+\frac{1}{2\left(1+\phi \right) }V\left( \phi \right)
\frac{1}{1+\phi }\left( \nabla
_{\varphi }\nabla ^{\varphi }-\square \right) \phi \notag
\\
+\frac{3}{2}\frac{1}{\phi (1+\phi )}\nabla _{\varphi } \phi \nabla ^{\varphi }\phi -\frac{3}{4}\frac{1}{\phi \left(1+\phi \right)}\nabla _{\lambda }\phi\nabla ^{\lambda }\phi ,\label{f3}
\end{eqnarray}
where there is no summation upon the pair of indices $\left(t,r,\theta ,\varphi \right)$.
A useful relationship is obtained by adding the gravitational field
equations Eqs.~(\ref{f1})--(\ref{f3}), from which we obtain the following
equation
\begin{multline}
e^{-\lambda }\left( \frac{\nu ^{\prime \prime }}{2}+\frac{\nu ^{\prime 2}}{4}
+\frac{\nu ^{\prime }}{r}-\frac{\nu ^{\prime }\lambda ^{\prime }}{4}\right)
=4\pi \frac{G}{1+\phi }\rho \left\langle u^{2}\right\rangle \\
+\frac{1}{1+\phi } V\left( \phi \right) +\frac{1}{1+\phi }\left( 2\nabla
_{t}\nabla ^{t}+\square \right) \phi -\frac{3}{\phi (1+\phi )}\nabla _t\phi \nabla ^t\phi,  \label{ff}
\end{multline}
where $\langle u^{2}\rangle =\langle u_{t}^{2}\rangle +\langle
u_{r}^{2}\rangle +\langle u_{\theta }^{2}\rangle +\langle u_{\varphi
}^{2}\rangle$.

Since we are interested in astrophysical applications at the extra-galactic
level, we may assume that the deviations from standard general relativity
(corresponding to the background value $\phi =0$) are small, i.e., $\phi \ll 1$. Thus, Eq.~(\ref{ff})
can be rewritten as
\be
e^{-\lambda }\left( \frac{\nu ^{\prime \prime }}{2}+\frac{\nu ^{\prime 2}}{4}
+\frac{\nu ^{\prime }}{r}-\frac{\nu ^{\prime }\lambda ^{\prime }}{4}\right)
\simeq 4\pi G\rho \left\langle u^{2}\right\rangle +4\pi G\rho _{\phi }^{(eff)},
\label{ff1}
\ee
where
\be\label{approx}
4\pi G\rho _{\phi }^{(eff)} \simeq  V\left( \phi \right) +\left( 2\nabla
_{t}\nabla ^{t}+\square \right) \phi -\frac{3}{\phi }\nabla _t\phi \nabla ^t\phi ,
\ee
represents an effective energy of the scalar field in the hybrid metric-Palatini gravitational model.

\section{The generalized virial theorem in hybrid metric-Palatini gravity}
\label{sec:c}

As a second step in the derivation of the virial theorem for galaxy clusters, which we
describe by using the distribution function $f_B$, we have to write down the
Boltzmann type equation governing the evolution of the distribution function.
This equation can then be integrated over the velocity space, to yield an
equation  which, used together with the gravitational field equations,
gives finally the required generalization of the virial theorem for a general relativistic distribution of point particles.

\subsection{The relativistic Boltzmann equation}

The transport equation for the distribution function  for a system of particles in a
curved arbitrary Riemannian space-time is given by the Boltzmann equation~\cite{Li66}
\be  \label{distr}
\left( p^{\alpha }\frac{\partial }{\partial x^{\alpha }}-p^{\alpha }p^{\beta
}\Gamma _{\alpha \beta }^{i}\frac{\partial }{\partial p^{i}}\right) f_B=0,
\ee
where $p^{\alpha }$ is the four-momentum of the particle, and $\Gamma _{\alpha \beta }^{i}$ are the Christoffel symbols associated to the metric. Note that the collissionless Boltzmann equation states that the local phase space density viewed by an observer co-moving with a star or galaxy is conserved.

A considerable simplification of the mathematical formalism can be obtained by introducing an appropriate
orthonormal frame or tetrad $e_{\mu }^{a}(x)$, $a=0,1,2,3$, which varies
smoothly over some coordinates neighborhood $U$ and satisfies the condition $%
g^{\mu\nu} e_{\mu}^{a} e_{\nu}^{b} =\eta^{ab}$ for all $x\in U$, where $\eta ^{ab}$ is the Minkowski metric tensor~\cite
{Li66,Ja70}. Any tangent vector $p^{\mu}$ at $x$ can be expressed as $%
p^{\mu}=p^{a}e_{a}^{\mu}$, which defines the tetrad components $p^{a}$.

In the case of the spherically symmetric line element given by Eq.~(\ref
{line}) an appropriate choice of the frame of orthonormal vectors is~\cite
{Li66,Ja70}:
\be
e_{\mu}^{0} = e^{\nu /2}\delta _{\mu}^{0}, \quad e_{\mu}^{1} =
e^{\lambda /2}\delta _{\mu }^{1}, \quad
e_{\mu}^{2} = r\delta_{\mu}^{2}, \quad  e_{\mu}^{3} = r\sin \theta
\delta_{\mu}^{3}.
\ee

Let $u^{\mu }$ be the four-velocity of a typical galaxy, satisfying the
condition $u^{\mu }u_{\mu }=-1$, with tetrad components $u^{a}=u^{\mu
}e_{\mu }^{a}$. In tetrad components the relativistic Boltzmann equation is
\begin{equation}
u^{a}e_{a}^{\mu}\frac{\partial f_B}{\partial x^{\mu}}+
\gamma_{bc}^{a}u^{b}u^{c}\frac{\partial f_B}{\partial u^{a}}=0,  \label{tetr}
\end{equation}
where the distribution function $f_B=f_B(x^{\mu },u^{a})$ and $\gamma
_{bc}^{a}=e_{\mu ;\nu }^{a}e_{b}^{\mu}e_{c}^{\nu}$ are the Ricci rotation
coefficients~\cite{Li66,Ja70}. By assuming that the only coordinate
dependence of the distribution function is upon the radial coordinate $r$,
Eq.~(\ref{tetr}) becomes~\cite{Ja70}
\begin{eqnarray}
u_{1}\frac{\partial f_B}{\partial r}- \left(\frac{1}{2}u_{0}^{2}\frac{%
\partial \nu}{\partial r}- \frac{u_{2}^{2}+u_{3}^{2}}{r}\right) \frac{%
\partial f_B}{\partial u_{1}}
    \nonumber \\
-\frac{1}{r}u_{1}\left( u_{2}\frac{\partial f_B}{\partial u_{2}} +u_{3}\frac{%
\partial f_B}{\partial u_{3}}\right)
    \nonumber \\
-\frac{1}{r}e^{\lambda /2}u_{3}\cot \theta \left( u_{2} \frac{\partial f_B}{%
\partial u_{3}}-u_{3} \frac{\partial f_B}{\partial u_{2}}\right) = 0.
\label{tetr1}
\end{eqnarray}


For a spherically symmetric system  the coefficient of $\cot \theta $ in Eq.~(\ref{tetr1}) must be zero. This implies that the distribution function $f_B$ is only a function of $r$, $u_{1}$ and $%
u_{2}^{2}+u_{3}^{2}$. Multiplying Eq.~(\ref{tetr1}) by $m u_{r} du $,
integrating over the velocity space, and by assuming that $f_B$
vanishes sufficiently rapidly as the velocities tend to $\pm \infty $, we
obtain
\bea
&&r\frac{\partial}{\partial r}\left[\rho\left\langle u_{1}^{2}\right\rangle%
\right]+ \frac{1}{2}\rho \left[ \left\langle u_{0}^{2}\right\rangle +
\left\langle u_{1}^{2}\right\rangle\right] r\frac{\partial \nu }{\partial r}
\nonumber\\
&&-\rho \left[ \left\langle u_{2}^{2}\right\rangle +\left\langle
u_{3}^{2}\right\rangle -2\left\langle u_{1}^{2}\right\rangle \right] =0.
\label{tetr2}
\eea

Multiplying Eq.~(\ref{tetr2}) by $4\pi r^{2}$, and integrating over the cluster gives~\cite
{Ja70}
\bea
&&\int_{0}^{R}4\pi \rho \left[ \left\langle u_{1}^{2}\right\rangle
+\left\langle u_{2}^{2}\right\rangle +\left\langle u_{3}^{2}\right\rangle%
\right] r^{2}dr \nonumber\\
&&-\frac{1}{2}\int_{0}^{R}4\pi r^{3}\rho \left[ \left\langle
u_{0}^{2}\right\rangle +\left\langle u_{1}^{2}\right\rangle\right] \frac{%
\partial \nu }{\partial r}dr=0.  \label{kin}
\eea

\subsection{Geometric quantities}

In the following we introduce some approximations that apply to test
particles in stable circular motion around galaxies, as well as  to the galactic
clusters. First, we assume that $\nu $ and $\lambda $ are slowly
varying functions of the radial coordinate (i.e.~$\nu^{\prime}$ and $\lambda^{\prime}$ are small). Therefore  in Eq.~(%
\ref{ff1}) we can neglect all the quadratic terms. Secondly, we assume that the motion of the galaxies is non-relativistic, so that
they  have velocities much smaller than the velocity of the light, i.e., $\langle u_{1}^{2}\rangle \approx \langle u_{2}^{2}\rangle \approx
\langle u_{3}^{2}\rangle \ll \langle u_{0}^{2}\rangle \approx 1$. Thus,
Eqs.~(\ref{ff1}) and (\ref{kin}) can be written as
\be  \label{fin1}
\frac{1}{2r^{2}}\frac{\partial }{\partial r}\left(r^{2} \frac{\partial \nu }{%
\partial r}\right) = 4\pi G\rho + 4\pi G\rho_{\phi}^{(eff)},
\ee
and
\be
2K-\frac{1}{2}\int_{0}^{R}4\pi r^{3}\rho \frac{\partial \nu }{\partial r}%
dr=0,  \label{cond1}
\ee
respectively, where
\begin{align}
K=\int_{0}^{R}2\pi \rho \left[ \left\langle u_{1}^{2}\right\rangle
+\left\langle u_{2}^{2}\right\rangle +\left\langle u_{3}^{2}\right\rangle %
\right] r^{2}dr,
\end{align}
is the total kinetic energy of the galaxies. The total baryonic mass of the system $M_B$ is
defined as $M_B=\int_{0}^{R}dM(r)=\int_{0}^{R} 4\pi \rho r^{2}dr$. The
main contribution to $M_B$ is due to the baryonic mass of the intra-cluster
gas and of the stars, but other particles, such as massive neutrinos, may
also give a significant contribution to $M_B$.

Now, multiplying Eq.~(\ref{fin1}) by $r^{2}$ and integrating from $0$ to $r$
we obtain
\begin{align}
GM_B(r)=\frac{1}{2}r^{2}\frac{\partial \nu }{\partial r}-GM_{\phi }^{(eff)}\left(
r\right),  \label{fin2}
\end{align}
where we have denoted
\begin{align}  \label{darkmass}
M_{\phi }^{(eff)}\left( r\right) =4\pi \int_{0}^{r}\rho _{\phi}^{(eff)}(r')r'^{2}
dr'.
\end{align}

Since in hybrid metric-Palatini gravity, the quantity $M_{\phi }^{(eff)}$ has essentially a geometric origin, we tentatively denote it as the \textit{geometric mass} of the
cluster.  In  the following we define the gravitational potential energies of the cluster as
\begin{eqnarray}
\Omega _B&=&-\int_{0}^{R}\frac{GM_B(r)}{r}\,dM_B(r), \\
\Omega _{\phi }^{(eff)}&=&\int_{0}^{R}\frac{GM_{\phi }^{(eff)}(r)}{r}\,dM_B(r),
\end{eqnarray}
respectively, where $R$ is the cluster radius.
By multiplying Eq.~(\ref{fin2}) with $dM_B(r)$, following an
integration from $0$ to $R$, we obtain the  relation
\begin{align}
\Omega _B=\Omega _{\phi }^{(eff)}-\frac{1}{2}\int_{0}^{R}4\pi r^{3}\rho \frac{\partial
\nu }{\partial r}\,dr.
\end{align}

\subsection{Generalized virial theorem}

Finally, with the use of Eq.~(\ref{cond1}), we arrive at the generalization of
the virial theorem in hybrid metric-Palatini gravity, which takes the familiar
form
\be
2K + \Omega = 0  \label{theor} \,,
\ee
where the total gravitational potential energy of the system, $\Omega$, defined as
\be
\Omega = \Omega _B- \Omega_{\phi }^{(eff)} \,, \label{theor2}
\ee
contains a contribution term consisting of a geometric origin, $\Omega_{\phi }^{(eff)} $.

The generalized virial theorem, given by Eq.~(\ref{theor}), can be
represented  in a more transparent physical  form if we introduce the radii $R_{V}$ and $%
R_{\phi }$, defined by
\be
R_{V}=M_B^{2}\Bigl/\int_{0}^{R}\frac{M_B(r)}{r}\,dM_B(r),\Bigr.
\ee
and
\be
R_{\phi }^{(eff)}=\left[M_{\phi }^{(eff)}\right]^{2}\Bigl/\int_{0}^{R}\frac{M_{\phi }^{(eff)}(r)}{r}\,dM_B(r),\Bigr.
\label{RU3}
\ee
respectively.  In analogy to the geometric mass considered above, the quantity $R_{\phi }$ may be denoted as the \textit{geometric radius} of the cluster of galaxies in the hybrid metric-Palatini gravitational models. Thus, the baryonic potential energy  $\Omega _B$ and the effective scalar field potential energy $\Omega _{\phi}^{(eff)}$ are finally given by
\begin{eqnarray}
\Omega _B &=&-\frac{GM_B^{2}}{R_{V}},
   \\
\Omega _{\phi}^{(eff)} &=&\frac{G\left[M_{\phi }^{(eff)}\right]^{2}}{R_{\phi }^{(eff)}},
\end{eqnarray}
respectively.

We define the virial mass $M_{V}$ of the cluster of galaxies as
\begin{align}
2K=\frac{GM_BM_{V}}{R_{V}}.
\end{align}

After substitution into the virial theorem, given by Eq.~(\ref{theor}), we
obtain the following relation between the virial and the baryonic mass of the galaxy cluster
\be  \label{fin6}
\frac{M_{V}}{M_B}=1+\frac{\left[M_{\phi }^{(eff)}\right]^{2}R_{V}}{M_B^{2}R_{\phi }^{(eff)}}.
\ee
If $M_{V}/M_B>3$, a condition which holds for most of the observed galactic
clusters, then Eq.~(\ref{fin6}) provides the virial mass in hybrid metric-Palatini gravity,
which can be approximated as
\be
M_{V}\approx \frac{\left[M_{\phi}^{(eff)}\right]^2}{M_B}\frac{R_{V}}{R_{\phi }^{(eff)}}.  \label{virial}
\ee

From the point of view of the astrophysical observations the virial mass $M_V$ is determined from the
study of the velocity dispersion $\sigma _r^2$ of the stars and of the galaxies in the clusters. According to the virial theorem in hybrid metric-Palatini gravity, most of the mass in a cluster with mass $M_{tot}$ is in the form of the geometric mass $M_{\phi }^{(eff)}$, so that $M_{\phi}^{(eff)} \approx M_{tot}$. An observational  possibility of detecting the presence of the geometric mass and of the astrophysical effects of hybrid metric-Palatini gravity is through gravitational lensing, which can provide direct evidence of the geometric mass distribution and of the gravitational effects associated to the presence of the scalar field even at distances extending far beyond of the virial radius of the galaxy cluster.

\section{Astrophysical applications}
\label{sec:d}

Once the integrated mass as a function of the radius is determined for
galaxy clusters, a physically meaningful fiducial radius for the mass
measurement has to be defined. The radii commonly used are either $r_{200}$
or $r_{500}$. These radii lie within the radii of the mean gravitational
mass density of the matter $\left\langle \rho _{tot}\right\rangle =200\rho
_{c}$ or $500\rho _{c}$, with $\rho_{c}$ given by $%
\rho_{c}\left( z\right) =h^{2}(z)3H_{0}^{2}/8\pi G$, where $h(z)$ is the
Hubble parameter normalized to its local value, i.e., $h^{2}(z)=\Omega
_{m}\left( 1+z\right) ^{3}+\Omega _{\Lambda }$, $\Omega _{m}$ is the
mass density parameter, and $\Omega _{\Lambda }$ is the dark energy density
parameter, respectively ~\cite{Ar05}. A pragmatic approach to the virial mass $M_{vir}$ is to use $%
r_{200}$ as the outer boundary of the galaxy cluster~\cite{ReBo02}. The numerical values of the
radius $r_{200}$ are in the range $r_{200}=0.85$ Mpc (for the cluster NGC
4636) and $r_{200}=4.49$ Mpc (for the cluster A2163), so that a typical
value for $r_{200}$ is approximately $2$ Mpc. The masses corresponding to $%
r_{200}$ and $r_{500}$ are denoted by $M_{200}$ and $M_{500}$, respectively,
and it is usually assumed that $M_{vir}=M_{200}$ and $R_{vir}=r_{200}$, where $R_{vir}$ is the virial radius of the cluster~\cite
{ReBo02}.

\subsection{The weak field approximation of hybrid metric-Palatini gravity}

In the limit of small static gravitational fields the metric tensor can be approximated as $g_{\mu \nu}\approx \eta _{\mu \nu}+h_{\mu \nu}$, where $\eta _{\mu \nu}$ is the Minkowski metric, and $h_{\mu \nu} \ll 1$. The local perturbation of the scalar field is denoted by $\varphi$, and therefore $\phi \approx \phi _0+\varphi $, where $\phi _0$ is the asymptotic value of the scalar field. In this approximation the evolution of the scalar field is described by the equation \cite{fX}
\be\label{eqi}
\left(\nabla ^2-\frac{1}{r_{\phi }^2}\right)\varphi =\kappa ^2\frac{\phi _0}{3}\rho _B,
\ee
 where $\rho _B$ is the mass density of the baryonic matter, $\kappa ^2=c^3/8\pi G$, and $r_{\phi }=1/\left(Gm_{\phi }/c^2\right)$, with the effective mass of the field given by $m_{\phi }^2=1/\left.\left[2V(\phi )-V'(\phi)-\phi(1+\phi)V''(\phi)/3\right]\right|_{\phi =\phi _0}$.

 The solution of Eq.~(\ref{eqi}) is given by
 \be
 \varphi (r)=\frac{2}{3}\phi _0\frac{GM_B}{c^2r}e^{-2r/r_{\varphi}},
 \ee
 where  $M_B=4\pi \int^R_0{\rho _Br^2dr}$ is the total baryonic mass. The scalar field potential can be written as $V(\phi)=V(\phi _0+\varphi)\approx V\left(\phi _0\right)+V'\left(\phi _0\right)\varphi$.

 In order to estimate the astrophysical effects of the scalar field we have to find first the explicit form of the effective energy associated to the scalar field $\rho _{\phi }^{(eff)}$, given by Eq.~(\ref{approx}). In the static case one can neglect all the derivatives with respect to the time, and take $\Delta \varphi =\varphi /r_{\phi }^2+\left(\kappa ^2\phi _0/3\right)\rho _B$.  Therefore for $\rho _{\phi }^{(eff)}$ we obtain
\bea\label{rhoeff}
 \frac{4\pi G}{c^4}\rho _{\phi }^{(eff)}(r)&\approx &V\left(\phi _0\right)+V'\left(\phi _0\right)\varphi+\frac{\varphi }{r_{\phi }^2}+\frac{\kappa ^2\phi _0}{3}\rho _B
 \nonumber\\
 & \approx &  V\left(\phi _0\right)+\frac{2}{3}\Phi _0\frac{GM_B}{c^2r}e^{-2r/r_{\varphi}}+\frac{\kappa ^2\phi _0}{3}\rho _B, \nonumber\\
\eea
 where $\Phi_0=V'\left(\phi _0\right)+1/r_{\varphi}^2$. The scalar field effective density is fixed in  terms of the scalar field potential, the Newtonian approximation of the scalar field, and the baryonic matter density, which naturally appears in the effective Klein-Gordon equation when taking the Newtonian limit of the model. In principle, $\rho _B$ should be obtained by solving the full set of gravitational field equations of the hybrid metric - Palatini gravity model, which can be done only by using numerical methods. Instead, in the following we adopt a phenomenological approach,  by assuming a simple inverse square functional form for the baryonic matter distribution in the cluster. Hence we  assume that the baryonic matter density inside the cluster has a $r^{-2}$ dependence on the distance from the cluster center, that is, $\rho _B\propto r^{-2}$, so that $\rho _B=\rho _{B0}r^{-2}$, with $\rho _{B0}$ a constant.

\subsection{Geometric mass and geometric radius from galactic cluster
observations}

In the clusters of galaxies most of the baryonic mass  is in the form of the
intra-cluster gas. The  gas mass density $\rho _{g}$  can be
fitted with the observational data by using the following simple expression~\cite{ReBo02}
\be
\rho _{g}(r)=\rho _{0}\left( 1+\frac{r^{2}}{r_{c}^{2}}\right) ^{-3\beta /2},
\label{dens}
\ee
where $r_{c}$ is the core radius, and $\rho _{0}$ and $\beta $ are
(cluster-dependent) constants.

One can also assume
that the pressure $P_{g}$ of the gas satisfies the ideal gas equation of state $%
P_{g}=(k_{B}T_{g}/\mu m_{p})\rho_{g}$, where $k_{B}$ is Boltzmann's
constant, $T_{g}$ is the gas temperature, $\mu \approx 0.61$ is the mean
atomic weight of the particles in the cluster gas, and $m_{p}$ is the proton
mass~\cite{ReBo02}. This assumption is well justified physically, since as shown by X-ray observations, the hot, ionized
intra-cluster gas is in isothermal equilibrium. Therefore, with the use of the Jeans equation~\cite{Bi87}
the total mass distribution of the matter in the cluster can be obtained as a function of the gas density as ~\cite{ReBo02,HaCh07}
\be
M_{tot}(r)=-\frac{k_{B}T_{g}}{\mu m_{p}G}r^{2}\frac{d}{dr}\ln \rho _{g}.
\ee

Taking into account the density profile of the gas given by Eq.~(\ref
{dens}), for the total mass profile inside the cluster we obtain the
 relation~\cite{ReBo02}
\be
M_{tot}(r)=\frac{3k_{B}\beta T_{g}}{\mu m_{p}G}\frac{r^{3}}{r_{c}^{2}+r^{2}}.
\label{mp}
\ee

According to the hybrid metric-Palatini gravitational model, the total mass $M_{tot}$ of
the cluster consists of the sum of the baryonic mass (mainly the
intra-cluster gas), and of the geometric mass, so that
\be
M_{tot}(r)=4\pi \int_{0}^{r}\left[ \rho _{g}+\rho _{\phi }^{(eff)}\right]
r^{2}dr.
\ee
 Therefore $M_{tot}(r)$ satisfies the
 mass continuity equation given by
\be
\frac{d }{dr}M_{tot}(r)=4\pi r^{2}\rho _{g}\left( r\right) +4\pi
r^{2}\rho _{\phi }^{(eff)}\left( r\right).  \label{massf}
\ee

With the use of Eqs.~(\ref{dens}) and (\ref{mp}) we
obtain the expression of the geometric density term inside the cluster as
\be
4\pi \rho _{\phi }^{(eff)}\left( r\right)=\frac{3k_{B}\beta T_{g}\left(
r^{2}+3r_{c}^{2}\right) }{\mu m_{p}\left( r_{c}^{2}+r^{2}\right) ^{2}}-\frac{%
4\pi G\rho _{0}}{\left( 1+r^{2}/r_{c}^{2}\right) ^{3\beta /2}}.
\ee
In the cluster region where  $r\gg r_{c}$ we obtain for $\rho _{\phi }^{(eff)}$ the simple relation
\be\label{rhoclust}
4\pi \rho _{\phi }^{(eff)}\left( r\right)=\left[ \frac{3k_{B}\beta T_{g}}{\mu m_{p}}%
-4\pi G\rho _{0}r_{c}^{3\beta }r^{2-3\beta }\right] \frac{1}{r^{2}}.
\ee

By power expanding the exponential in Eq.~(\ref{rhoeff}) we obtain
\bea\label{rhotheor}
4\pi \rho _{\phi }^{(eff)}(r)&\approx &\frac{c^4}{G}\Bigg[
 V\left(\phi _0\right)-\frac{4}{3}\frac{GM_B}{c^2r_{\varphi}}\Phi _0\nonumber\\
 &&+\frac{2}{3}\Phi _0\frac{GM_B}{c^2r}+\frac{\kappa ^2\phi _0}{3}\frac{\rho _{B0}}{r^2}\Bigg].
\eea
In the following we neglect the constant terms in the effective geometric density, as corresponding to the cosmological background. Therefore the comparison of Eqs.~(\ref{rhoclust}) and (\ref{rhotheor}) gives
\be\label{60}
\beta =\frac{1}{3},
\quad
\phi _0=\frac{3}{8\pi \rho _{B0}}\frac{k_BT_g}{\mu m_p},
\quad
\Phi_0=-\frac{6\pi G}{c^2}\frac{\rho _0r_c}{M_B}.
\ee

Thus, from the last condition, i.e., $\phi_0 < 0$, the scalar field potential must satisfy the condition
\be
V'\left(\phi _0\right)+1/r_{\varphi }^2<0.
\ee
A possible potential satisfying this condition is the exponential potential $V(\phi)=V_0\exp (-2\alpha \phi)$, with $V_0$ and $\alpha $ positive constants, which gives $V'\left(\phi _0\right)=-2\alpha \exp\left(-2\alpha \phi _0\right)<0$. Therefore for this case $r_{\varphi}$ must satisfy the condition  $r_{\varphi}>(1/2\alpha)\exp\left(2\alpha \phi _0\right)$.

The geometric mass can be obtained, by using the observational data, as
\bea
&& GM_{\phi}^{(eff)}(r) = 4\pi\int_{0}^{r} r^{2} \rho _{\phi}^{(eff)}(r)\,dr
\nonumber\\
&& \hspace{-1cm} =  \frac{3k_{B}\beta T_{g}}{\mu m_{p}} \frac{r}{1+r_{c}^{2}/r^{2}} -4\pi G \rho_{0}\int_{0}^{r}
  \frac{r^{2}dr}{\left(1+r^{2}/r_{c}^{2}\right)^{3\beta/2}},
\eea
and in the limit $r\gg r_{c}$ it can be approximated as
\begin{align}
GM_{\phi }^{(eff)}\left( r\right) \approx \left[ \frac{3k_{B}\beta T_{g}}{\mu m_{p}}-%
\frac{4\pi G\rho _{0}r_{c}^{3\beta }r^{2-3\beta }}{3\left( 1-\beta \right) }%
\right] r.  \label{GM}
\end{align}

By using the approximation given by Eq.~(\ref{rhotheor}) for the geometric density, we obtain
\be\label{64}
 M_{\phi}^{(eff)}(r)\approx \frac{c^4}{G}\left(\frac{1}{3}\Phi _0\frac{GM_B}{c^2}r^2+\frac{\kappa ^2\phi _0}{3}\rho _{B0}\,r\right).
 \ee

 Eqs.~(\ref{60}) and (\ref{64}) show that the problem of the equivalent description of dark matter in hybrid metric - Palatini  gravity has a self-consistent solution, and that the free parameters of the model can be determined from observations. The obtained values can be tested observationally, at least in principle, by using Eq.~(\ref{64}).

 Astronomical observations \cite{ReBo02} suggest that the value of the parameter $\beta $, appearing in the gas density profile, given by Eq.~(\ref{dens}) is of the order of $\beta \approx 2/3$. Let's assume now for $\beta $  value of the order of $\beta\approx 2/3$, or higher. In this case, the observational effective ``dark matter'' density, profile given by Eq.~(\ref{rhoclust}),  goes like $4\pi \rho _{\phi }^{(eff)}\left( r\right)\approx {\rm constant}/ r^2$. On the other hand,  the Newtonian limit of the hybrid metric-Palatini gravity predicts a theoretical ``dark matter'' profile of the form $4\pi \rho _{\phi }^{(eff)}\left( r\right)\approx a_1 / r + a_2 / r^2$, where $a_1$ and $a_2$ are model-dependent constants.  To be in agreement with Eq.~(\ref{rhoclust}), favored by the observations, requiring $\beta \approx 2/3$, we just need to set $C_1=0$, or very small. This is equivalent to setting $\Phi_0=V'\left(\phi _0\right)+1/r_{\varphi}^2\approx 0$, which is just a condition imposed on the self-interaction potential $V(\varphi )$ of the scalar field. An exact solution to this condition always exists, and, therefore, our theoretical model can also reproduce the effective observational ``dark matter'' density profile, given by Eq.~(\ref{rhoclust}), when the more realistic $\beta=2/3$ value is adopted.

\subsection{Radial velocity dispersion in galactic clusters}

In terms of the characteristic
velocity dispersion $\sigma _{1}$ the virial mass can  also be expressed  as~\cite{Ca97}
\be
M_{vir}=\frac{3}{G}\sigma _{1}^{2}R_{vir}.
\ee
Taking into account that the velocity distribution in the cluster is isotropic, we
have $\langle u^2\rangle=\langle u_1^2\rangle+\langle u_{2}^2\rangle+\langle
u_{3}^2\rangle=3\langle u_1^2\rangle=3\sigma _r^2$, where $\sigma _r^2$ is
the radial velocity dispersion. $\sigma _1$ and $\sigma _r $ are related by $%
3\sigma _1^2=\sigma _r^2$.

The radial velocity dispersion relation for clusters of
galaxies in hybrid metric-Palatini gravity can be derived from Eq.~(\ref{tetr2}). Since the velocity distribution is isotropic, we obtain first
\be
\frac{d}{dr}\left( \rho \sigma _{r}^{2}\right) +\frac{1}{2}\rho \frac{d\nu }{%
dr}=0.  \label{veldis}
\ee

Inside the cluster in the first order of approximation the condition $e^{-\lambda}\approx 1$ holds.
In the limit of small velocities
the modified field equation, Eq.~(\ref{ff1}) can be integrated to yield
\be\label{58}
r^{2}\nu ^{\prime }=\frac{2GM_{\phi }(r)}{c^2}+\frac{2GM_B(r)}{c^2}+2C,
\ee
where $C$ is an arbitrary constant of integration. At the boundary of the cluster, where $r=R_{vir}$, the metric of the spherically symmetric matter distribution can be taken as approximately Schwarzschild, with $\nu \approx \ln\left(1-2GM_{vir}/c^2R_{vir}\right)$, giving $\left.r^2\nu '\right|_{r=R_{vir}}\approx 2GM_{vir}/c^2\approx 2GM_{\phi }/c^2$.  By estimating Eq.~(\ref{58}) at $r=R_{vir}$ gives $C=GM_B\left(R_{vir}\right)/c^2$.

Since from Eq.~(\ref{veldis}) we have $\nu ^{\prime }=-(2/\rho )d\left( \rho \sigma_{r}^{2}\right) /dr$, it follows that in hybrid metric-Palatini gravity the radial velocity dispersion of the galactic clusters satisfies the differential equation
\begin{align}
\frac{d}{dr}\left( \rho \sigma _{r}^{2}\right) =-\frac{GM_{\phi }(r)}{c^2r^{2}}%
\rho (r)-\frac{GM_B(r)}{c^2r^{2}}\rho (r)-\frac{C}{r^{2}}\rho (r),
\end{align}
with the general solution given by
\bea
\sigma _{r}^{2}(r)&=&-\frac{1}{\rho }\int^{r} \biggl[ \frac{GM_{\phi
}(r')}{c^2r'^{2}}\rho (r')
\nonumber\\
&&
+\frac{GM_B(r')}{c^2r'^{2}}\rho (r')
+\frac{C}{r'^{2}}\rho (r')\biggr] dr'+\frac{C_{1}}{%
\rho },  \label{veldisf}
\eea
where $C_{1}$ is an integration constant.

In the following we consider a simple case in which the density $\rho $ of the normal matter inside the cluster has a
power law distribution, given by
\be
\rho (r)=\rho _B(r)=\rho _{0\gamma }r^{-\gamma },
\ee
with $\rho _{0\gamma }$ and $\gamma \neq 1,3$ positive constants. The corresponding
baryonic matter mass profile is $M_B(r)=4\pi \rho _{0\gamma }r^{3-\gamma }/\left(
3-\gamma \right) $. For the geometric mass we assume that it is given by Eq.~(\ref{64}).

Therefore, for $\gamma \neq 1,3$, we obtain the following expression for the velocity dispersion,
\bea
\sigma _{r}^{2}(r)&=&-\frac{1}{3}\frac{\Phi _0GM_B}{1-\gamma}r+\frac{\kappa ^2\phi _0c^2\rho _{B0}}{3\gamma}+\nonumber\\
&&\frac{2\pi G\rho _{0\gamma }}{\left(
\gamma -1\right) \left( 3-\gamma \right) }r^{2-\gamma }+\frac{C}{\gamma +1}%
\frac{1}{r}+\frac{C_{1}}{\rho _{0\gamma }}r^{\gamma }.
\eea
For  $\gamma =1$, we find
\bea
\sigma _{r}^{2}(r)&=&-\frac{1}{3}\Phi _0GM_Br\ln r+\frac{\kappa ^2\phi _0c^2\rho _{B0}}{3}+\nonumber\\
&&\frac{C}{2r}-2\pi G\rho _{0\gamma }r\ln r+\frac{C_{1}}{%
\rho _{0}}r,
\eea
while for $\gamma =3$ we obtain
\bea
\sigma _{r}^{2}(r)&=&\frac{\Phi _0GM_B}{6}r+\frac{\kappa ^2\phi _0c^2}{3}-\nonumber\\
&&\pi G\rho _{0}\left( \ln r+\frac{1}{4}
\right) \frac{1}{r^{4}}+\frac{C}{4}\frac{1}{r}+\frac{C_{1}}{\rho _{0}} r^{3}.
\eea

The numerical value of the integration constant $C_1$ can be determined from the knowledge of the radial velocity dispersion relation $\sigma _r^2(r)$ at some   radius $r_0$.

For clusters of
galaxies the observed data for the velocity dispersion  are usually analyzed by assuming for the radial velocity dispersion the simple form $\sigma
_{r}^{2}(r)=B/(r+b) $, with $B$
and $b$ constants. For the density of the galaxies in the
clusters the profile $\rho \left( r\right) =A/r\left( r+a\right)
^{2}$, with $A$ and $a$ constants, is used. The observational data are then
fitted with these functions by using a non-linear fitting
procedure~\cite{Ca97}. For $r\ll a$, $\rho (r)\approx A/r$, while
for $r\gg a$, $\rho (r)$ behaves like $\rho (r)\approx A/r^{3}$.
Therefore the comparison of the observed velocity dispersion
profiles of the galaxy clusters and the velocity dispersion
profiles predicted by hybrid metric-Palatini gravity provides
a powerful method for observationally discriminating between the different
modified gravity theoretical models.

\section{Discussions and final remarks}

\label{sec:e}

Modern astrophysics and cosmology have to face
two major difficulties: the dark energy and the dark
matter problems, respectively. One promising approach  to
improve our understanding of these issues is to modify gravity at large galactic and cosmological scales. In
particular, the hybrid metric-Palatini gravitational theory  can challenge the need
for dark matter and  dark energy. In the framework of this theory, cosmological models that
account for the late time acceleration of the universe do exist \cite{fX1}, as well as viable
models that pass all the required Solar System tests \cite{fX}.

Moreover, as shown in the present paper, in the framework of hybrid metric-Palatini-type modified theories of gravity, the galactic dynamics of massive test particles may also be understood without the need for dark matter.
We have analyzed the ``dark matter'' problem by
considering a generalized version of the virial theorem. The virial
theorem was obtained by using a method based on the collisionless
Boltzmann equation. The additional scalar field  terms present in the
modified gravitational field equations give an effective
contribution to the gravitational energy, which at the
galactic/extra-galactic level acts as an effective mass, playing
the role of the ``dark matter". The total virial mass of the
galactic clusters is mainly determined by the effective mass
associated to the new scalar field terms, which can be called geometrical mass, since its intrinsic origin is geometrical. The presence of this  term may  explain the well-known virial theorem mass discrepancy in clusters of galaxies \cite{Bi87}.

In the framework of the hybrid metric-Palatini gravitational theory we have also shown the existence of
a strict proportionality between the virial mass of the cluster and its baryonic mass, a relation which can
also be tested observationally.

One of the important, and observationally testable,  predictions of the hybrid metric-Palatini gravitational ``dark matter'' model is that the geometric masses associated to the clusters, as well as its gravitational  effects, extend beyond the virial radii of the clusters. Observationally, the virial mass $M_V$ is obtained from the study of the velocity dispersions of the stars in the cluster. Due to the observational uncertainties, this method  cannot give a reliable estimation of the numerical value of the total mass $M_B+M_{\phi }^{(eff)}$ in the cluster. However, a much more powerful method for the determination of the total mass distribution in clusters is the gravitational lensing of light, which may provide direct evidence for the gravitational effects at large distances from the cluster, and for the existence of the geometric mass. The presence of hybrid metric-Palatini modified gravity effects at large distances from the cluster, and especially the large extension of the geometric mass, may lead to significantly different lensing observational signatures, as compared to the standard relativistic/dark matter model case. The bending angle in the hybrid metric-Palatini gravity models could be larger
than the one predicted by the standard dark matter models. Therefore, the detailed observational study of the gravitational lensing could discriminate between the different theoretical models introduced  to
explain the motion of galaxies (``particles") in the clusters of galaxies, and the standard dark matter models.

To conclude, the generalized virial theorem in hybrid metric-Palatini gravity is an efficient tool in observationally testing the viability of this class of generalized gravity models.

\acknowledgments S. C. is supported by INFN (Iniziativa Specifica NA12). T. S. K. is supported by the Research Council of Norway.  F. S. N. L. acknowledges financial support of the Funda\c{c}\~{a}o para a Ci\^{e}ncia e Tecnologia through the grants CERN/FP/123615/2011 and CERN/FP/123618/2011. G. J. O. is supported by the Spanish grants FIS2008-06078-C03-02, FIS2011-29813-C02-02, the Consolider Programme CPAN (CSD2007-00042), and the JAE-doc program of the Spanish Research Council (CSIC).


\begin{thebibliography}{99}
\bibitem{Bi87}  J.~Binney and S.~Tremaine, Galactic dynamics, Princeton,
Princeton University Press, (1987); M.~Persic, P.~Salucci and F.~Stel,
Month.\ Not.\ R.\ Astron. Soc. \textbf{281}, 27 (1996); A.~Borriello and
P.~Salucci, Month.\ Not.\ R.\ Astron. Soc. \textbf{323}, 285 (2001);
P.~Salucci, A.~Lapi, C.~Tonini, G.~Gentile, I.~Yegorova and U.~Klein,
astro-ph/0703115 (2007).

\bibitem{Sal} P. Salucci, C. Frigerio Martins, and A. Lapi, arXiv:1102.1184 (2011).

\bibitem{OvWe04}  J.~M.~Overduin and P.~S.~Wesson, Phys.\ Repts.\ \textbf{402%
}, 267 (2004); A. S. Majumdar and N. Mukherjee, Int. J. Mod. Phys. D14, 1095 (2005).

\bibitem{Harko:2011nu}
  T.~Harko and F.~S.~N.~Lobo,
  Phys.\ Rev.\ D {\bf 83}, 124051 (2011);
  %
  T.~Harko and F.~S.~N.~Lobo,
  Astropart.\ Phys.\  {\bf 35}, 547 (2012).

\bibitem{dark}  M.~Milgrom, Astrophys.\ J.\ \textbf{270}, 365 (1983);
R.~H.~Sanders, Astron.\ Astrophys.\ \textbf{136}, L21 (1984); J.~Bekenstein
and M.~Milgrom, Astrophys.\ J.\ \textbf{286}, 7 (1984); R.~H.~Sanders,
Astron.\ Astrophys.\ \textbf{154}, 135 (1986); P.~D.~Mannheim, Astrophys.\
J.\ \textbf{419}, 150 (1993); J.~W.~Moffat and I.~Y.~Sokolov, Phys.\ Lett.\
\textbf{B378}, 59 (1996); P.~D.~Mannheim, Astrophys.\ J.\ \textbf{479}, 659
(1997); M.~Milgrom, New Astron.\ Rev.\ \textbf{46}, 741, (2002); M.~Milgrom,
Astrophys.\ J.\ \textbf{599}, L25, (2003); M.~K.~Mak and T.~Harko, Phys.\
Rev.\ \textbf{D70}, 024010 (2004); J.~D.~Bekenstein, Phys.\ Rev.\ \textbf{D70%
}, 083509 (2004); M.~D.~Roberts, Gen.\ Rel.\ Grav.\ \textbf{36}, 2423,
(2004); J.~R.~Brownstein and J.~W.~Moffat, Astrophys.\ J.\ \textbf{636}, 721
(2006); J.~R.~Brownstein and J.~W.~Moffat, Mon.\ Not.\ Roy.\ Astron.\ Soc.\
\textbf{367}, 527 (2006); T.~Harko and K.~S.~Cheng, Astrophys.\ J.\ \textbf{%
636}, 8 (2006); C.~G.~B\"ohmer and T.~Harko, Class.\ Quantum Grav.\ \textbf{%
24}, 3191 (2007); S. Capozziello,  V.F. Cardone, A. Troisi, Mon. \ Not. \ Roy. \ Astron. \ Soc. \textbf{375}, 1423  (2007);  Bertolami, C. G. Boehmer, T. Harko, and F. S.N. Lobo, 	Phys. Rev. {\bf D75}, 104016 (2007); C. G. Boehmer, T. Harko, and F. S. N. Lobo, Astropart. Phys. {\bf 29}, 386 (2008);
S. Capozziello, E. De Filippis, and V. Salzano,  Mon.\ Not. \ Roy. \ Astron. \ Soc. \textbf{394}, 947  (2009);   L. A. Gergely, T. Harko, M. Dwornik, G. Kupi, and Z. Keresztes, Mon. Not. Royal Astron. Soc. {\bf 415}, 3275 (2011);  S. Capozziello and M. De Laurentis, Phys. \ Rept.  \textbf{509},   167 (2011);  K. C. Wong, T. Harko, K. S. Cheng, and L. A. Gergely, Phys. Rev. {\bf D86}, 044038 (2012); D. F. Mota, V. Salzano, S. Capozziello, and N. R. Napolitano, 	arXiv:1211.1019 (2012);
S. Capozziello and M. De Laurentis, Annalen Phys. {\bf 524}, 545  (2012).

\bibitem{Col} G. W. Collins, The virial theorem in stellar astrophysics, Tucson, Ariz., Pachart Publishing House (1978).

\bibitem{vir}  S.~Bonazzola, Astrophys.\ J.\ \textbf{182}, 335 (1973);
C.~Vilain, Astrophys.\ J.\ \textbf{227}, 307 (1979); A.~Georgiou, J.\ Phys.\
A: Math.\ Gen.\ \textbf{13}, 3751 (1980); E.~Gourgoulhon and S.~Bonazzola,
Class.\ Quantum Grav.\ \textbf{11}, 443 (1994); A.~Georgiou, Class.\ Quantum
Grav.\ \textbf{20}, 359 (2003).

\bibitem{Ja70}  J.~C.~Jackson, Month.\ Not.\ R.\ Astr.\ Soc.\ \textbf{148},
249 (1970).

\bibitem{No}  M.~Nowakowski, J.-C.~Sanabria and A.~Garcia, Phys.\ Rev.\
\textbf{D66}, 023003 (2002); A.~Balaguera-Antol\'{i}nez, C.~G.~B\"ohmer and
M.~Nowakowski, Class.\ Quantum Grav.\ \textbf{23}, 485 (2006);
A.~Balaguera-Antol\'{i}nez, D.~F.~Mota and M.~Nowakowski, Mon. Not. R. Astron. Soc. {\bf 382}, 621 (2007); M. Roshan, Class. Quantum Grav. {\bf 29}  215001 (2012); P. Mach, Mon. Not. R. Astron. Soc. {\bf 422}, 772 (2012).

\bibitem{HaCh07}  T.~Harko and K.~S.~Cheng, Phys. Rev. \textbf{D76}, 044013
(2007).

\bibitem{BoHaLo08} C. G. Boehmer, T. Harko, and F. S. N. Lobo, JCAP {\bf 0803}, 024 (2008).

\bibitem{Shah} H. R. Sepangi and S. Shahidi,  Class. Quant.  Grav. {\bf 26}, 185010 (2009); 	
M. Heydari-Fard and M. Heydari-Fard, Phys. Rev. {\bf D84}, 024040 (2011).

\bibitem{Sep} A. S. Sefiedgar, K. Atazadeh, and H. R. Sepangi, Phys. Rev. {\bf D80}, 064010 (2009).

\bibitem{fX} T. Harko, T. S. Koivisto, F. S. N. Lobo, and G. J. Olmo, Phys. Rev. {\bf D85}, 084016 (2012).

\bibitem{fX1} S. Capozziello, T. Harko, T. S. Koivisto, F. S.N. Lobo, and G. J. Olmo,  arXiv:1209.2895 (2012) to appear in Phys. Rev. D.

\bibitem{fX2} S. Capozziello, T. Harko, T. S. Koivisto, F. S.N. Lobo, and G. J. Olmo,  arXiv:1209.5862 (2012).

\bibitem{Ar05}  M.~Arnaud, ``X-ray observations of clusters of galaxies'',
in ``Background Microwave Radiation and Intracluster Cosmology'',
Proceedings of the International School of Physics ``Enrico Fermi'', edited
by F.~Melchiorri and Y.~Rephaeli, published by IOS Press, The Netherlands,
and Societ\`{a} Italiana di Fisica, Bologna, Italy, p.77, (2005).

\bibitem{ReBo02}  T.~H.~Reiprich and H.~B\"oringer, Astrophys.\ J.\ \textbf{%
567}, 716 (2002).

\bibitem{Sch01}  P.~Schuecker, H.~B\"ohringer, K.~Arzner and T.~H.~Reiprich,
Astron.\ Astrophys.\ \textbf{370}, 715 (2001); M. Baldi, V. Pettorino, G. Robbers, and V. Springel,  Mon. Not. R. Astron. Soc. {\bf  403}, 1684 (2010); 2B. Li and J. D. Barrow, Phys. Rev. {\bf D83}, 024007 (2011).

\bibitem{Li66}  R.~W.~Lindquist, Annals of Physics \textbf{37}, 487 (1966);
R.~Maartens and S.~D.~Maharaj, J.\ Math.\ Phys.\ \textbf{26}, 2869 (1985);
S.~Bildhauer, Class. Quantum Grav. \textbf{6}, 1171 (1989); Z.~Banach and
S.~Piekarski, J.\ Math.\ Phys.\ \textbf{35}, 4809 (1994).

\bibitem{NFW} J. F. Navarro, S. C. Frenk, and S. D. White,  The Astrophysical Journal {\bf 463},  563 (1996).


\bibitem{Ca97}  R.~G.~Carlberg, H.~K.~C.~Yee and E.~Ellingson, Astrophys.\
J.\ \textbf{478}, 462 (1997).

\end{thebibliography}
\end{document}